\documentclass{emulateapj}%[preprint2]{aastex6}
\usepackage{amsmath,amsgen,amstext,amsbsy,amsopn,amsfonts}
\usepackage{url,graphicx,tabularx,array}
\usepackage{listings}

\usepackage{color}
\definecolor{red}{rgb}{0.7,0.1,0.1}

\definecolor{green}{rgb}{0.1,0.7,0.1}

\definecolor{blue}{rgb}{0.2,0.2,0.8}

\begin{document}
\title{Assessing Astrophysical Uncertainties in Direct Detection with Galaxy Simulations}
\author{Jonathan D. Sloane,\altaffilmark{1} Matthew R. Buckley,\altaffilmark{1} Alyson M. Brooks,\altaffilmark{1} and Fabio Governato\altaffilmark{2}}
\affiliation{$^1$Department of Physics and Astronomy, Rutgers University, Piscataway, NJ 08854, USA}
\affiliation{$^2$Astronomy Department, University of Washington, Box 351580, Seattle, WA 98195-1580}

\begin{abstract}
    We study the local dark matter velocity distribution in simulated Milky Way-mass galaxies, generated at high resolution with both dark matter and baryons. We find that the dark matter in the Solar neighborhood is influenced appreciably by the inclusion of baryons, increasing the speed of dark matter particles compared to dark matter-only simulations. The gravitational potential due to the presence of a baryonic disk increases the amount of high velocity dark matter, resulting in velocity distributions which are more similar to the Maxwellian Standard Halo Model than predicted from dark matter-only simulations. Further, the velocity structures present in baryonic simulations possess a greater diversity than expected from dark matter-only simulation. We show the impact on the direct detection experiments LUX, DAMA/Libra, and CoGeNT using our simulated velocity distributions, and explore how resolution and halo mass within the Milky Way's estimated mass range impact the results.  A Maxwellian fit to the velocity distribution tends to overpredict the amount of dark matter in the high velocity tail, even with baryons, and thus leads to overly optimistic direct detection bounds on models which are dependent on this region of phase space for an experimental signal. Our work further demonstrates that it is critical to transform simulated velocity distributions to the lab frame of reference, due to the fact that velocity structure in the Solar neighborhood appears when baryons are included.  There is more velocity structure present when baryons are included than in dark matter-only simulations. Even when baryons are included, the importance of the velocity structure is not as apparent in the Galactic frame of reference as in the Earth frame.
\end{abstract}


\section{Introduction}
Dark matter is a key ingredient in understanding cosmology, galaxy formation and galaxy evolution.
The addition of non-baryonic dark matter is essential for a quantitative understanding of the Cosmic Microwave Background \citep{Smoot1992}, large scale structure \citep{Davis1985}, galaxy interactions \citep{Clowe2006}, and galaxy kinematics \citep{Rubin1980}.
Dark matter is also evidence of physics beyond the Standard Model of particle physics, as no known particle has the requisite properties to be dark matter. 
If dark matter is a relic of thermal production in the early Universe, a confluence of scales makes a weakly interacting massive particle (WIMP) in the 10-1000 GeV mass range an attractive candidate ({\it i.e.}, the WIMP ``Miracle''), but a WIMP is by no means the only possibility \citep{Steigman1985}.

Probing the new physics of dark matter can be done in many ways. For dark matter with WIMP-like interactions with Standard Model particles, these methods include collider searches, indirect detection, and direct detection. Direct detection experiments, the focus of this work, search for dark matter-nucleon interactions resulting in measurable nuclear recoils in a low-background target material. The interpretations of the experimental results in terms of dark matter-Standard Model interactions are subject to astrophysical uncertainties in the local dark matter density and velocity distributions. Such uncertainties need to be controlled to allow comparison between direct detection and other classes of dark matter experiments.

In this work, we aim to quantify the astrophysical uncertainties on direct detection results. We examine in detail the changes in the interpretation of results from LUX \citep{FirstLUXResults}, CoGeNT \citep{CoGeNT2013}, and DAMA/Libra \citep{DAMALIBRA2010} due to different dark matter velocity distributions that can arise from a range of merger histories for Milky Way-mass galaxies.  LUX currently places the strongest upper limits on dark matter direct detection, but both CoGeNT and DAMA/Libra have anomalous results which have been claimed to be evidence of dark matter. The negative results from LUX place a great deal of pressure on a dark matter interpretation of the CoGeNT and DAMA/Libra signals. However, these experiments are sensitive to very different regions of the dark matter velocity distribution, and it has been suggested that the addition of baryons to galaxy simulations could alter the velocity distribution in such a way that it might eliminate the tension between experiments \citep[e.g.,][]{Ling2010}.  We explore that possibility in this paper.

Different  direct detection experiments probe different portions of the velocity distribution due to different detector energy thresholds and different target nucleon masses. Heavier target nuclei are  sensitive only to higher impact velocities, assuming equal detector thresholds. Therefore, to convert the experimental results into a limit (or preferred region) on the dark matter-nucleon scattering cross section, some assumption of the dark matter velocity distribution must be made. Currently, all direct detection collaborations employ a Maxwell-Boltzmann velocity distribution (in the Galactic reference frame), known as the Standard Halo Model (SHM).
This is acknowledged as a convenient, but not realistic, benchmark using a distribution that dates to before simulations were capable of creating realistic galaxies.
$N$-body simulations have shown significant deviations from Maxwell-Boltzmann \citep{Vogelsberger2009, Kuhlen2010, Ling2010, Lisanti2011, Pillepich2014, Butsky2015}, and new fitting functions are still being proposed to  accommodate non-Maxwellian velocity distributions. However, these updated fitting functions are either based on a single simulated halo that includes baryonic (gas and stellar) physics \citep{Lisanti2011}, or on simulations that do not include baryonic physics at all \citep{Mao2013I}.

Large suites of fully cosmological simulations are required to find new fitting functions. Such suites exist almost exclusively as dark matter-only simulations, due to the prohibitive computational expense required to run a suite with baryons.  The suites of simulations which result in large numbers of galaxies including baryonic effects are generally too low resolution to accurately simulate internal galaxy kinematics/dynamics.  In this work, rather than search for a new fitting function, we use a small suite of high resolution simulations to further investigate the impact of baryonic physics on the velocity distribution of dark matter in the solar neighborhood, for a range of galaxy merger histories.  Work by \citealt{Pillepich2014} examined the difference between the Milky Way analogue Eris and ErisDark \citep{Guedes2011} to determine how baryonic effects can change the dark matter velocity distribution for a single galaxy.
We complement and extend \citealt{Pillepich2014} by presenting Milky Way analogue galaxies at similar resolution 
to Eris but spanning a range of merger histories. 

Resolution is important in correctly modeling galaxy formation. High resolution simulations allow a more realistic prescription for star formation and feedback, which has been shown to influence the dark matter distribution through potential fluctuations due to rapid gas relocation \citep{Mashchenko2008, Governato2010, Governato2012, Pontzen2012, Martizzi2013, Pontzen2014, DiCintio2014}.
It has recently become possible to simulate the evolution of a galaxy in a fully cosmological context, including baryonic effects, with  $\lesssim$150 pc resolution. The simulations used in this work produce realistic rotation curves \citep{Christensen2014}, match the stellar mass -- metallicity relationship \citep{Brooks2007, Christensen2015}, the size -- luminosity relation \citep{Brooks2011}, 
produce gas outflows that lead to realistic angular momentum distributions, including low-mass bulgeless galaxies \citep{Brook2011, Brooks&Christensen2016}, reproduce observational trends in gas fraction, and match the the stellar mass to halo mass relation at z=0 \citep{Munshi2013}. 

Note that our simulation of the dark matter local velocity is independent of the particle nature of dark matter (barring extremely large dark matter self-interactions), as long as the dark matter is cold. Thus, our results have potential implications in a number of search strategies for dark matter, not just direct detection searches for WIMP-like dark matter.  For example, the ADMX experiment \citep{Duffy:2006aa,Asztalos:2009yp} searches for axion dark matter via axion-photon conversion in a strong magnetic field. The line shape for the expected signal depends on the assumption for the dark matter velocity distribution. A more accurate understanding of this distribution could allow for more powerful constraints from this experiment.

In Section \ref{Simulations} we describe the simulations used in more detail, and present results of their dark matter velocity distribution in Section \ref{VelDist}, before moving to the effects on direct detection in Section \ref{DirectDetection}.
As this paper was being prepared, two papers \citep{Bozorgnia:2016ogo,Kelso:2016qqj} performed similar analyses of the effect of baryonic physics on the dark matter velocity distribution in Milky Way-like galaxies. While we agree in some respects with their conclusions, most notably in finding an increase in the amount of high-velocity dark matter when baryons are included in the simulations, we disagree on other important points. We discuss possible sources of discrepancy in Section \ref{Discussion}.
We conclude that using the SHM generally results in overly optimistic constraints at low dark matter masses, but the shift in experimental constraints caused by our more accurate velocity distributions does not alleviate the tension between the LUX null-results and the signals seen by DAMA/Libra and CoGeNT. Regardless, in order to accurately compare the constraints on dark matter parameters set by different classes of dark matter experiments, it is imperative that experimenters consider baryonic effects on the resulting dark matter velocity distribution within galaxies.  We summarize our conclusions in Section \ref{Conclusions}.

%%%%%%%%%%%%%%%%%%%%%%%%%%%%%%%%%%%%%%%%%%
\section{Simulations}\label{Simulations} %
%%%%%%%%%%%%%%%%%%%%%%%%%%%%%%%%%%%%%%%%%%
We separate our analysis into two separate groups of galaxies, one used to explore the role of baryons on the dark matter velocity distribution in the Solar neighborhood, and the other used to explore the role of resolution and halo mass.  We discuss the galaxies used in our mass/resolution tests in section~\ref{ResSims}.  

Advances in computational power as well as algorithmic and parallel implementation improvements have led to the ability to simulate galaxies in a fully cosmological context with force resolution only a few 100 pc or less \citep[e.g.,][]{Brook2012, Aumer2013, FIRE2014, Christensen2015}. 
Cold collisionless dark matter works very well in accounting for large scale structure, but on small scales where non-gravitational effects become important dark matter-only simulations do poorly in matching observed properties of galaxies \citep[for a review, see][]{Brooks2014}. 
 To simulate the galaxies that reside in dark matter halos requires modeling the gas physics, for which we use the method of smooth particle hydrodynamics (SPH). SPH discretizes the gas into sample points which are used to approximate the fluid quantities. 
The primary galaxies used in this work were simulated with the $N$-body + SPH code \textsc{Gasoline} \citep{Wadsley2004}.

We will first show results for four galaxies (designated h239, h258, h277 and h285) that span a halo mass range of $0.7-0.9 \times 10^{12} M_\odot$. This is on the lower half of the Milky Way-mass range $0.6 - 1.3 \times 10^{12}M_{\odot}$ \citep{Kafle2012}. 
These galaxies are drawn from a simulation box of 50~Mpc on a side and resimulated using the `zoom-in' technique \citep{Katz1993}. This allows us to focus resolution on the region of interest, $\sim$1 Mpc centered on the galaxy, while still keeping the gravitational effects of large scale structure. Each galaxy has two versions run from the same initial conditions: a dark matter-only run and an $N$-body + SPH run. The cosmology used is based on  {Wilkinson Microwave Anisotropy Probe (WMAP)} year 3 parameters: $\Omega_M = 0.26,~\Omega_\Lambda = 0.74,~h=0.73,~\sigma_8 = 0.77,~n=0.96$ \citep{Spergel2007}.  {We discuss the role of cosmology further in Section \ref{MAHSection}.} Star formation is modeled by following the creation and destruction of H$_2$, and only allowing stars to form in H$_2$, as in \citealt{Christensen2012}.  Star particles are born with a Kroupa Initial Mass Function \citep{Kroupa1993}. Stellar feedback uses the blastwave method as detailed in \citealt{Stinson2006}, and includes metal line cooling as in \citealt{Shen2010MetalLineCooling}. These galaxies were run with a spline gravitational force softening equivalent to 174~pc in the high resolution region.  The mass of the dark matter particles in the dark matter-only runs is $1.8\times10^5$ M$_\odot$. The SPH galaxies have a dark matter particle mass that is lower by a factor of (1$ - f_b$), where $f_b$ is the cosmic baryon ratio, $\Omega_b/\Omega_m$, and is 0.175 for the adopted cosmology, i.e., $1.5\times10^5$ M$_\odot$. Gas particles begin with a mass of $2.7\times10^4$ M$_\odot$, and stars are born with $30\%$ of the parent gas particle's mass.  

It has been shown that the inclusion of baryonic physics circularizes the galaxy's potential \citep[e.g.,][]{Butsky2015, Bryan2013}, and leads to higher central concentrations \citep{Oman2015}. Furthermore baryons aid in the destruction of substructure by increasing the central potential and therefore the ability to tidally disrupt accreted structures \citep{Penarrubia2010,Brooks&Zolotov2014}. The disk also effectively shreds any substructure on an orbit through it \citep{Ostriker1972, DOnghia2010}.  The sphericalization of the potential and the addition of destroyed substructure can both contribute to the change in dark matter velocity distribution between a dark matter-only run and a run that includes baryonic physics.

Each galaxy has a qualitatively different merger history: h239 was continually bombarded with small galaxies, h258 had a $1:1$ major merger at $z\sim1$, h277 had its last major merger at $z\sim3$, and h285 underwent a five way merger at $z\sim1.7$ followed by a $1:12$ mass ratio counter-rotational merger beginning at $z\sim1.4$ and ending at $z\sim0.8$ that seems to have altered its structural properties (see below). All end up as Milky Way-mass galaxies at the present day.

Simulation h258 has a dark disk; this has been studied in lower resolution versions of this run  \citep{Read2009} and is still present in the higher resolution simulation used here. With its relatively quiescent merger history, h277 is thought to have the most similar merger history to the Milky Way.  Galaxy h227 does not show evidence for a dark disk, in agreement with observations of the Milky Way \citep[e.g.,][]{Ruchti2015}. \citet{Loebman2014} noted that h277's structural properties (e.g., disk scale length, bulge-to-disk ratio, maximum circular velocity) are within 10\% of the Milky Way's. The galaxy h285 has a counter-rotating component in the bulge, likely related to the counter rotating merger mentioned earlier that finishes at $z=0.8$.  {Table \ref{SimTable}} shows each galaxy's virial mass \footnote{As discussed in \citet{Munshi2013} and \citet{Sawala2012}, SPH halo masses are generally lower than the same halo in a dark matter-only run by $\sim 5-10$\%. This is attributed to feedback.  At Milky Way masses, this is primarily due to the fact that feedback removes material, and thus fitting to the same overdensity leads to a slightly smaller virial radius.  For the one galaxy in Table \ref{SimTable} in which the SPH run appears more massive, it is due to infalling substructure in the SPH case that isn't yet infalling in the dark matter-only run.}, the escape velocity from the Solar neighborhood, and the number of dark matter particles in the Solar neighborhood (defined in section~\ref{VelDist}).

%%%%%%%%%%%%%%%%%%%%%%%%%%%%
%Simulation Parameter Table% Search for: Parameter Table Belongs Here-ish
%%%%%%%%%%%%%%%%%%%%%%%%%%%%
%\capstartfalse
\begin{deluxetable*}{lcccccc}
\tablecaption{Relevant Simulated Galaxy Properties\label{SimTable}} 
\tablecolumns{6}
\tablewidth{\linewidth}
\tablehead{\colhead{Simulation} & \colhead{$M_{\text{halo}}~[10^{12} M_\odot]$}
   & \colhead{$v_{\text{esc}}$~[km s$^{-1}$]} & \colhead{$v_0$~[km s$^{-1}$]} & \colhead{z @ 20\%} & \colhead{z @ 50\%} & \colhead{$N_{\rm dm}$}}
\startdata
h239      & 0.91 & 471 & 204 & 1.8 & 1.1 & 5847 \\	
h239Dark  & 0.93 & 411 & 155 & 1.9 & 1.1 & 3424 \\
h258      & 0.77 & 458 & 185 & 2.3 & 1.3 & 6482 \\
h258Dark  & 0.82 & 403 & 152 & 2.3 & 1.2 & 2878 \\
h277      & 0.68 & 462 & 190 & 3.4 & 2.2 & 7460 \\
h277Dark  & 0.74 & 389 & 148 & 3.1 & 1.6 & 3327 \\
h285      & 0.88 & 474 & 187 & 2.2 & 1.7 & 7115 \\
h285Dark  & 0.73 & 378 & 135 & 2.2 & 2.0 & 2887 
\enddata

\tablecomments{Details of the four simulated Milky Way-like galaxies considered in this paper: dark matter halo mass, escape velocity from the Solar neighborhood, best fit Maxwellian's peak velocity,  {the redshift at which 20\% of the final mass was assembled, the redshift at which 50\% of the final mass was assembled,} and number of dark matter halo particles in the annulus defining the Solar region.  Each galaxy was simulated both with and without baryons -- the latter are designated ``Dark'' in the table.}
\end{deluxetable*}
%\capstarttrue

\subsection{Mass/Resolution Tests}\label{ResSims}

In Section~\ref{Discussion}, we use two additional galaxies that span the Milky Way's estimated mass range, to examine the effects of mass and resolution. To do this, we use two additional galaxies (h329 and h148) that bracket a mass range similar to the uncertainty in the Milky Way's mass.  Both h329 and h148 were run with the same resolution (174 pc force resolution) as the four galaxies presented above, but h148 was also run at twice the force resolution (87 pc) and 8 times higher mass resolution. These galaxies were run to $z = 1$ using {\sc Gasoline}'s successor ChaNGa \citep{Menon2015}, and have a Planck cosmology \citep{Planck}. In addition to the physics described above, the baryonic runs include black holes and black hole feedback, and 50\% greater supernova feedback.  The black holes necessitate an increase in the dark matter mass resolution, which is now $10^3$ M$_{\odot}$. The black hole implementation is described in detail in \citet{Tremmel2015}.

The same properties listed in  {Table~\ref{SimTable}} are listed for these new galaxies in  {Table~\ref{AppSimTable}}. The high resolution version of h148 is listed as h148Hi.  Both h148 and h148Hi have also been run as dark matter-only. 

\section{Dark Matter Velocity Distributions}\label{VelDist}

\begin{figure*}[!t]
\centering
\includegraphics[width=2.\columnwidth]{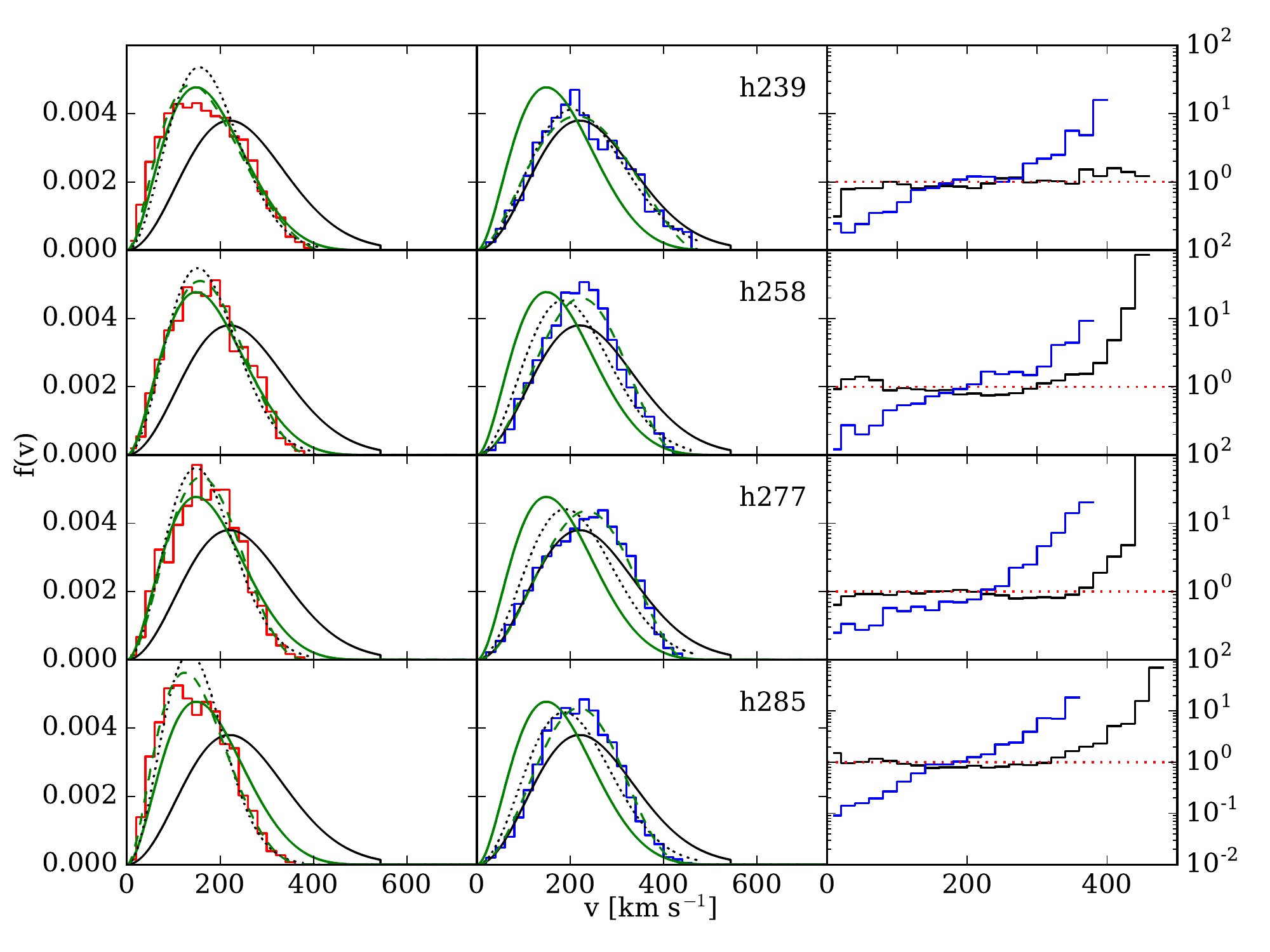}
\caption{Local dark matter velocity distribution in the Galactic reference frame. Left: Dark matter-only runs (red). Center: Baryonic runs (blue).  Each halo is compared to the SHM (solid black) and the best empirical fit from \citealt{Mao2013I} (solid green). A fit to the individual halos using the Mao parameterization (green dashed line) and a Maxwellian (black dotted line) are also shown. Right: Ratio of SPH to dark matter-only $f(v)$ (blue line), and the ratio of the SHM to SPH $f(v)$ (black line), in the Galactic reference frame.  {The red dotted line is unity.}}

\label{f_galactic}
\end{figure*}

Only the dark matter distribution at the Earth's neighborhood is relevant for direct detection, but the best cosmological galaxy simulations have resolution limits on the order of 100 parsecs. The Sun's distance from the Galactic Center also has significant uncertainties, \citep[e.g.,][]{Bovy2012} and so we must average over a resolved region to provide an expected averaged velocity distribution. We have chosen to average the velocity distribution over a cylindrical annulus in the plane of the disk. The height and width of the annulus are 1 kpc, and the central radius is 8 kpc.
We have oriented the annulus in the dark matter-only simulations to coincide with the annulus in the SPH simulations, centered on the minimum of the gravitational potential in each simulation\footnote{We also explored defining the disk in the dark matter-only run to be a plane normal to the halo's angular momentum vector.  However, in three of the runs the angular difference between this plane and the baryonic disk was less than 6 degrees.  In the fourth, the vector was nearly flipped 180 degrees.  Hence, using this alternative definition does not impact our conclusions.}. 
It is the high resolution of our simulations, allowing finer sampling with more particles, that enables us to look at such a relatively small region.

%%%%%%%%%%%%%%%%%%%%%%%%%%%%%%%%%%%%%%
%Auxillary Simulation Parameter Table%
%%%%%%%%%%%%%%%%%%%%%%%%%%%%%%%%%%%%%%
%\capstartfalse
\begin{deluxetable*}{lcccccc}
\tablecaption{Simulated Galaxy Properties for Resolution/Mass Tests\label{AppSimTable}} 
\tablecolumns{6}
\tablewidth{\linewidth}
\tablehead{\colhead{Simulation} & \colhead{$M_{\text{halo}}~[10^{12} M_\odot]$}
   & \colhead{$v_{\text{esc}}$~[km s$^{-1}$]} & \colhead{$v_0$~[km s$^{-1}$]} & \colhead{z @ 20\%} & \colhead{z @ 50\%} & \colhead{$N_{\rm dm}$}}
\startdata
h329        & 0.46 & 446 & 180 & 3.0 & 1.4 & 20049 \\
h148        & 1.10 & 603 & 254 & 1.9 & 0.8 & 31696 \\
h148Dark    & 1.06 & 469 & 173 & 1.9 & 0.8 & 3282  \\
h148Hi      & 1.12 & 585 & 230 & 1.9 & 0.8 & 69198 \\
h148HiDark  & 1.06 & 469 & 171 & 1.9 & 0.8 & 26473
\enddata

\tablecomments{Details of the various versions of the two simulated Milky Way-like galaxies used to consider the impact of mass and resolution. Columns are the same as in  {Table~\ref{SimTable}}. The versions designated ``Dark'' in the table include only dark matter. These properties are at redshift 1.}
\end{deluxetable*}
%\capstarttrue

For both the dark matter-only and baryonic versions of the simulations, the local dark matter velocity distribution in the galactic reference frame (shown in Figure \ref{f_galactic}) already displays clear differences from the Maxwellian distribution assumed by experiments. The Maxwellian often used, the SHM, assumes a spherical, isothermal halo with $v_0=220$ km s$^{-1}$ and $v_{\rm esc}=544$ km s$^{-1}$:
\[
f(v) \propto
\begin{cases}
	v^2 \exp{\left(-v/v_0\right)^2}, & \text{if } |v| \in [0,v_{\rm esc}] \\
    0, & \text{otherwise}
\end{cases}
\]
We show this SHM in Figure~\ref{f_galactic} as a black solid line, and is the same in all panels. The black dotted lines show the best fit Maxwellian to each simulation (using $v_0$ and $v_{esc}$ from Table 1). 
In Figure~\ref{f_galactic}, we also show (green solid line) the empirical velocity distribution from \citet{Mao2013I}:
\[
f(v) \propto 
\begin{cases}
    \exp \left(-\frac{|v|}{v_M} \right) \left( v_{\rm esc}^2 - |v|^2\right)^p, & \text{if } |v| \in [0,v_{\rm esc}] \\
    0, & \text{otherwise}
\end{cases}
\]
where $v$ is the dark matter velocity, $v_{\rm esc}$ is escape velocity from the region under consideration and $v_M$ and $p$ are parameters that are extracted from fits to dark matter-only simulations. We have adopted their best fit parameters for Milky Way-mass galaxies in the solid green line, which is the same in every panel.  The dashed green lines in each panel of Figure~\ref{f_galactic} show the \citet{Mao2013I} parameterization, but with parameters fit to our individual halos.  

\begin{figure*}[!t]
\centering
\includegraphics[width=2\columnwidth]{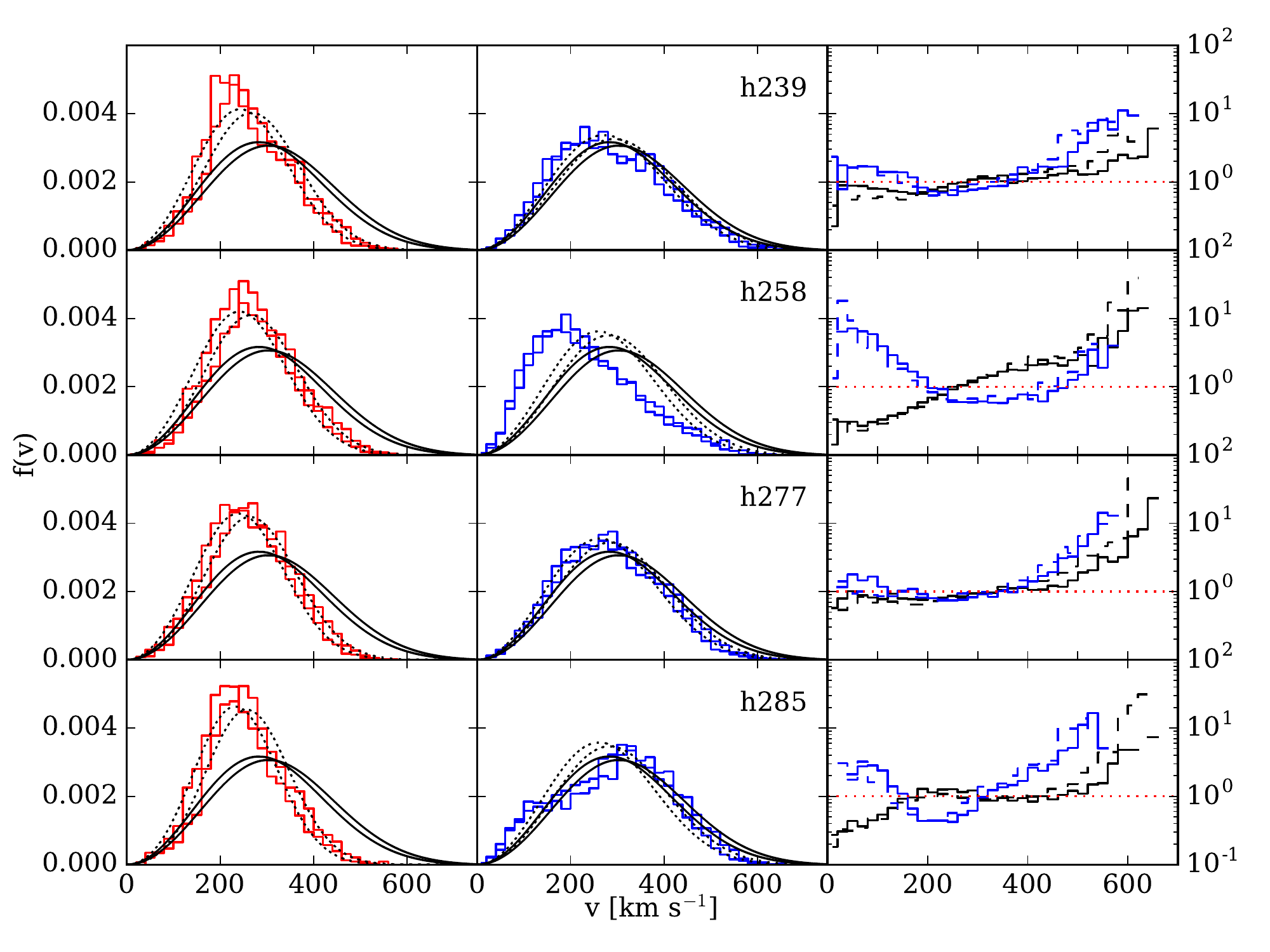}
\caption{Velocity distributions in the Earth reference frame at days of highest and lowest flux. Left: Dark matter-only runs (red). Center: Baryonic runs (blue). In all cases, we compare the distributions to the SHM (black). 
The {  Maxwellian} fits to our individual halos {  (black dotted lines)} are boosted from the Galactic rest frame shown in Figure~\ref{f_galactic}.  They are not re-fitted in the Earth frame.  {The highest flux day corresponds to the distribution that reaches higher velocities.} Right: Ratio of SPH to dark matter-only $f(v)$ (blue lines), and the ratio of the SHM to SPH $f(v)$ (black lines), in the Earth reference frame. The solid line is the ratio at the day of  maximum flux, and the dashed line is the ratio at the day of minimum flux.  {The red dotted line is unity.}}
\label{f_earth} 
\end{figure*}

It can be seen in Figure~\ref{f_galactic} that there is less high velocity material in all of the dark matter-only and baryonic runs than in the SHM. This same effect was seen in the study by \citet{Kuhlen2010}, where they examined halos on the more massive end of the Milky Way's allowed mass, demonstrating that this deficit remains even for more massive dark matter-only halos. We examine the role that halo mass plays in the high velocity tail in Section~\ref{Discussion}. 

Importantly, the baryonic runs have more high velocity dark matter than the dark matter-only runs.  This is expected due to the increased potential in the disk because of the presence of baryons. Because of the added disk potential, the SHM is a better fit to the SPH simulations than the dark matter-only halos.\footnote{In fact, \citet{Butsky2015} suggested that a Gaussian is a better fit to the dark matter velocity distribution in runs with baryons.  While this also seems true in some of our galaxies, a Gaussian isn't always a good fit.}  
That the SHM is a better fit in the baryonic simulations becomes more apparent in the Earth reference frame.  The dark matter velocity distribution in the Earth reference frame, shown in Figure \ref{f_earth}, at the days of maximum and minimum flux, is built from the particle velocities in the Galactic reference frame with a Galilean boost to the Solar reference frame, azimuthally averaging the simulation data.
The transformation to the Earth frame is then carried out on a particle by particle basis (creating a different distribution for each day). The Sun-Earth velocities used in the transformation are as in \citet{AnnualModulationGelminiGondolo2001} and \citet{FreeseLisantiSavage2012}, with one exception: the local standard of rest was changed to make our analysis more consistent with the stellar kinematics of the simulated galaxies. Rather than use the relative speed of the Milky Way's local standard of rest, 235 km s$^{-1}$, we use the peak of the simulated local stellar velocity distribution, typically about 195 km s$^{-1}$ (built from star particles in the same region from which the dark matter was selected).  We apply this value, derived in the baryonic run, to the corresponding dark matter-only run.  We also boost the SHM by 195 km s$^{-1}$ instead of the more typical 235 km s$^{-1}$.

For 3 of our 4 galaxies, the SHM does a better job fitting the SPH runs in the Earth reference frame than in the Galactic reference frame  (note that we shift the fit from the Galactic reference frame; we do not try a new fit in the Earth frame).  {This appears to be due to the boost. The Galactic reference frame adopts the {\it magnitude} of the velocity of all of the dark matter particles in the Solar annulus that we defined,  but the Earth reference frame accounts for directionality. The transformation of the SHM and the Maxwellian fit into the Solar/Earth reference frames assumes isotropy in velocity space.  On the other hand, the simulation distributions are boosted on a particle by particle basis, taking into account the components of the velocity for each particle. While our dark matter particle distributions are largely isotropic in the dark matter-only runs, our baryonic simulations instead consistently enhance the dark matter velocities in a preferred direction, usually the direction of net spin of the stellar disk.  When boosted, this directionality appears to a higher velocity tail more consistent with the SHM in 3 of our 4 galaxies.  In the other galaxy it leads to lower velocities.  The result depends on the particular velocity structure in any given halo. }

There are several sources which add preferred direction to the dark matter velocities in the baryonic simulations.    
Inclusion of baryons has been found to leave a halo's total spin parameter relatively unchanged, but can affect the spin parameter of the inner halo by as much as 30-50\% \citep{Bryan2013, Bett2010}.  The halo spin is separate from a dark disk which can add further coherent structure. Some of the galaxies studied here have unambiguous dark matter co-rotation with the stellar disk, evidence of a dark disk.  One also shows enhanced counter-rotation with respect to the stellar disk. Additionally, \citet{Kuhlen2012} and \citet{Vogelsberger2009} used dark matter-only simulations to show that debris flow can influence the velocity structure.  It remains to be examined whether the inclusion of baryons enhances this debris flow structure further, though more destruction of satellites is expected in the presence of a baryonic disk than found in dark matter-only simulations \citep{Brooks2013}.

Using these simulated velocity distributions, we now turn to the effect on direct detection searches for dark matter scattering with nucleons.

%%%%%%%%%%%%%%%%%%%%%%%%%%%%%%%%%%%%%%%%%%%%%%%%%%%%%%%%%%%%%%%%%%%%%%%%%%%%%%%

\section{Direct Detection}\label{DirectDetection}

Direct detection searches for dark matter have restricted the available WIMP parameter space considerably in the previous decade. As is usually done, we will examine the relation between nucleon cross section $\sigma_N$ and the dark matter particle mass $m_\chi$. The robustness of this restriction hinges on a good understanding of the systematic uncertainties, of which the astrophysical uncertainty is an important (and relatively unconstrained) part. In order to accurately compare results of collaborations using different targets and detectors, as well as comparing to the collider and indirect detection searches, these uncertainties need to be quantified. Below, we show the effect on exclusion limits of using dark matter velocity distributions from our simulations.

\begin{figure*}[!t]
\centering
\includegraphics[width=2\columnwidth]{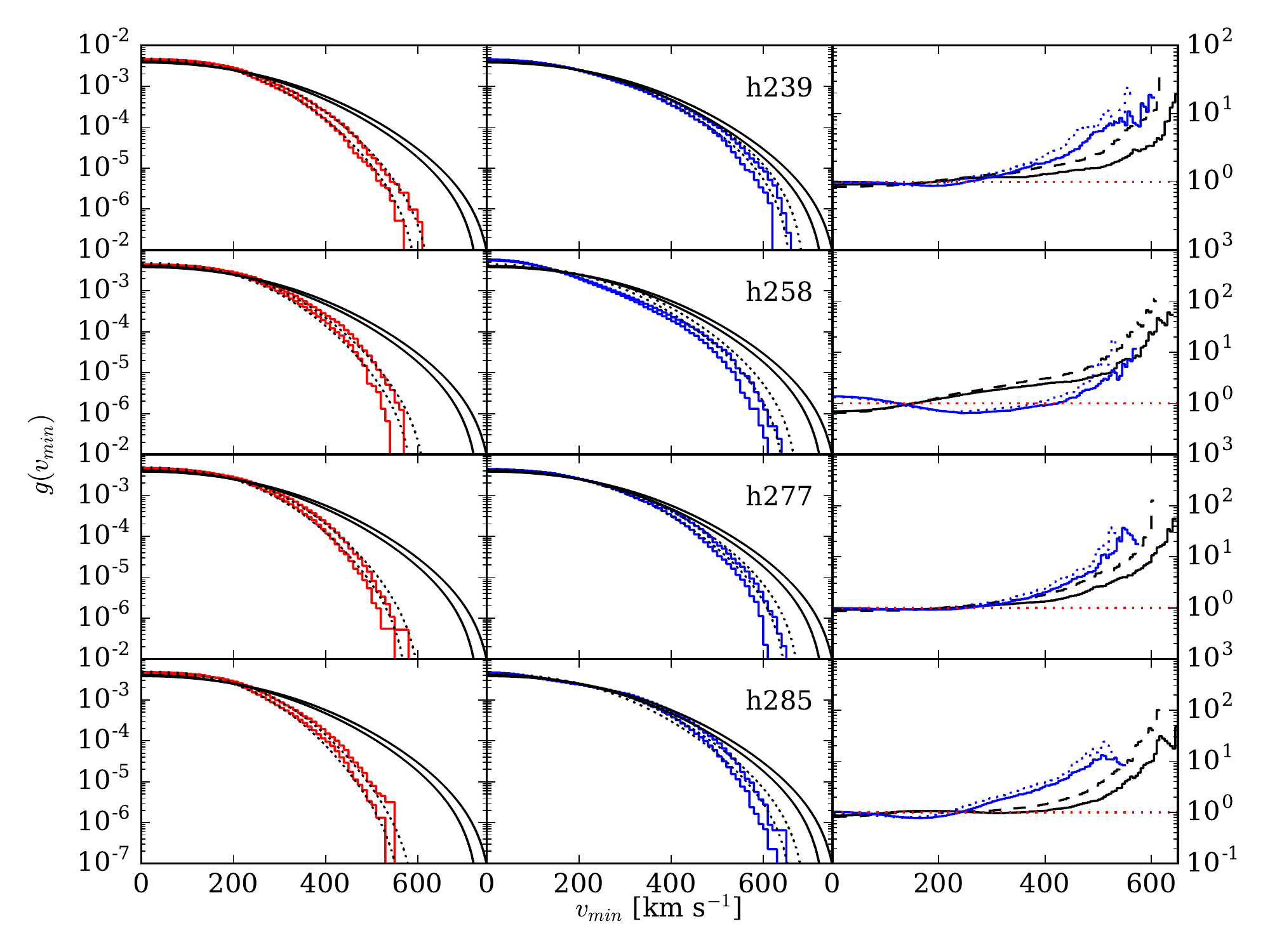}
\caption{Plots of the weighted velocity distribution integral $g(v_{\rm min})$ in the Earth reference frame at the days of highest and lowest flux. Left: dark matter-only simulations (red).  Center: Baryonic simulations (blue). We show also the SHM (solid black) and best fit Maxwellian to individual halos (dotted black).  {The highest flux day corresponds to the distribution that reaches higher velocities.} Right: Ratio of SPH to dark matter-only $g(v_{\rm min})$ (blue lines), and the ratio of the SHM to SPH $g(v_{\rm min})$ (black lines), in the Earth reference frame. The solid line is the ratio at the day of maximum flux, and the dotted/dashed line is the ratio at the day of minimum flux. The SHM consistently overpredicts the fraction of high-velocity dark matter in the baryonic runs, and underpredicts the low-velocity.  {The red dotted line is unity.}}
\label{g_earth}
\end{figure*}

The rate of dark matter interactions is given by:
\begin{align}\label{detectionRate}
    \frac{dR}{dE_R}(E_R) = \frac{d\sigma}{dE_R} \frac{\rho_0}{m_\chi} \int_{v_{\rm min}}^{\infty}\frac{f(v)}{v} dv 
\end{align}
with $\rho_0$ the local dark matter density, 
$v_{\rm esc}$ the escape velocity from the Solar neighborhood, $v_{\rm min} \equiv \sqrt{E_R m_T/(2 \mu_{T\chi}^2)}$ is the minimum velocity recoil a particular detector can measure. This is dependent on the nucleon recoil energy $E_R$, the nucleus target mass $m_T$, and the nucleus-dark matter reduced mass $\mu_{T\chi}$. 
In addition to the local dark matter density, the rate of dark matter events is 
proportional to the integrated weighted velocity function 
$g(v_{\rm min}) \equiv \int_{v_{\rm min}}^\infty \frac{f(v)}{v} dv $. This quantity is plotted in Figure \ref{g_earth} in the Earth rest frame for each of our considered distributions. As each experiment has a different $v_{\rm min}$ to which it is sensitive, each experiment probes  {different regions of $f(v)$}. Thus, while the degree of agreement between different experiments can be quantified in a relatively astrophysics-independent manner \citep{Fox2011,DelNobile2013}, one's confidence in the extracted limits on the more fundamental dark matter parameters which enter into $d\sigma/dE_R$ is limited by the uncertainties in $f(v)$ and $g(v_{\rm min})$. 

There is some tension between anomalous positive results from several direct detection experiments, and the null results of other experiments which should be sensitive to the putative signals. 
The strongest limits are set by the LUX Collaboration \citep{FirstLUXResults}, with many other collaborations setting competitive and complimentary limits.\footnote{
As this paper was being completed, LUX released an updated analysis with a larger data set and improved analysis techniques for the low-mass dark matter region \cite{Akerib:2015rjg}. To calculate the effect of our simulated velocity distributions on these new results would require knowledge of the detector response under the new analysis. This has not yet been made public. Therefore, we continue to use the LUX data from \citep{FirstLUXResults} in this work. The trends we observe can be reasonably extrapolated to the newest LUX bounds.}
 Heavy WIMP dark matter ($m_\chi \gtrsim 100$ GeV) has much stronger exclusion bounds from LUX when compared to the bounds on light dark matter, ($m_\chi \sim10$ GeV). The CoGeNT Collaboration \citep{CoGeNT2013} has seen events above their background expectation and the DAMA/Libra Collaboration \citep{DAMALIBRA2010} has observed an exceptionally strong modulation signal, each of which can be interpreted as originating from dark matter collisions, though such conclusions are in extreme tension with the non-observation of signal by LUX.

These collaborations, like all others, assume the SHM in their analyses. As we have seen, this does not agree with dark matter velocity distributions found in simulations. As these collaborations are sensitive to different portions of the dark matter velocity distribution due to different target materials and detector sensitivities, it is not immediately clear if the tension between the positive and negative results may be a result of a non-Maxwellian velocity distribution. This has been addressed in the literature \citep{Fox2011,DelNobile2013} by changing to a different parameter space, $\eta$ vs.~$v_\text{min}$, that clearly shows which regions of velocity-space different experiments are sensitive to, assuming a particular dark matter mass. The $\eta$ parameter, equivalent to $g(v_\text{min})$ up to normalization factors, defined as:
\begin{align}
    \eta(v_{\rm min}) \equiv \frac{ \rho_0 \sigma }{m_\chi} \int_{v_{\text{min}}} \frac{f(v)}{v} dv
\end{align}
is approximately independent of the detector response \citep{Fox2011} and is therefore directly comparable across experiments, without relying on a particular assumption for $f(v)$. These studies find it is difficult (but perhaps not impossible) for the CoGeNT, DAMA/Libra, and LUX results to all be consistent with an arbitrary function $\eta(v_{\rm min})$.  As previously mentioned, while this technique allows for more direct comparison between direct detection experiments, it does not allow for extraction of the cross section $\sigma$, which would allow for comparison between direct detection experiments and other types of dark matter searches ({\it e.g.}~collider searches and indirect detection experiments). This motivates our efforts to determine $f(v)$ directly from simulation.

The interaction rate Eq.~\eqref{detectionRate} depends on the local dark matter density only as an overall multiplicative factor. As such it effects each experiment in the same way.
Given that direct detection collaborations generally use $\rho_0 \sim 0.3 - 0.4$ GeV/cm$^3$ as the local dark matter density, this cannot contribute to positive signals in one experiment overlapping excluded parameter space in another.
Work by \citet{LisantiSpergel2012} has further shown that it is unlikely that the Sun is in a local overdensity of dark matter, and so using a canonical value of $\rho_0\sim 0.3-0.4$  GeV/cm$^3$ is indeed reasonable. Older sources generally use $\rho_0\sim 0.3$ GeV/cm$^3$, while newer use $\rho_0\sim 0.4$ GeV/cm$^3$ from \citet{Pato2015}, but this change amounts to a simple rescaling of experimental results.
Here, we adopt $0.4$ GeV/cm$^3$ to allow comparison with current literature. Our simulations all have $\rho_0\sim 0.3-0.4$  GeV/cm$^3$, despite the presence of a dark disk in some of our simulated galaxies. We leave the exploration of the role of the dark disk for future work, but, because our simulations yield a consistent dark matter density, the relevant astrophysical source of uncertainty is the dark matter velocity distribution.

\begin{figure*}[!t]
\centering
\includegraphics[width=0.8\textwidth]{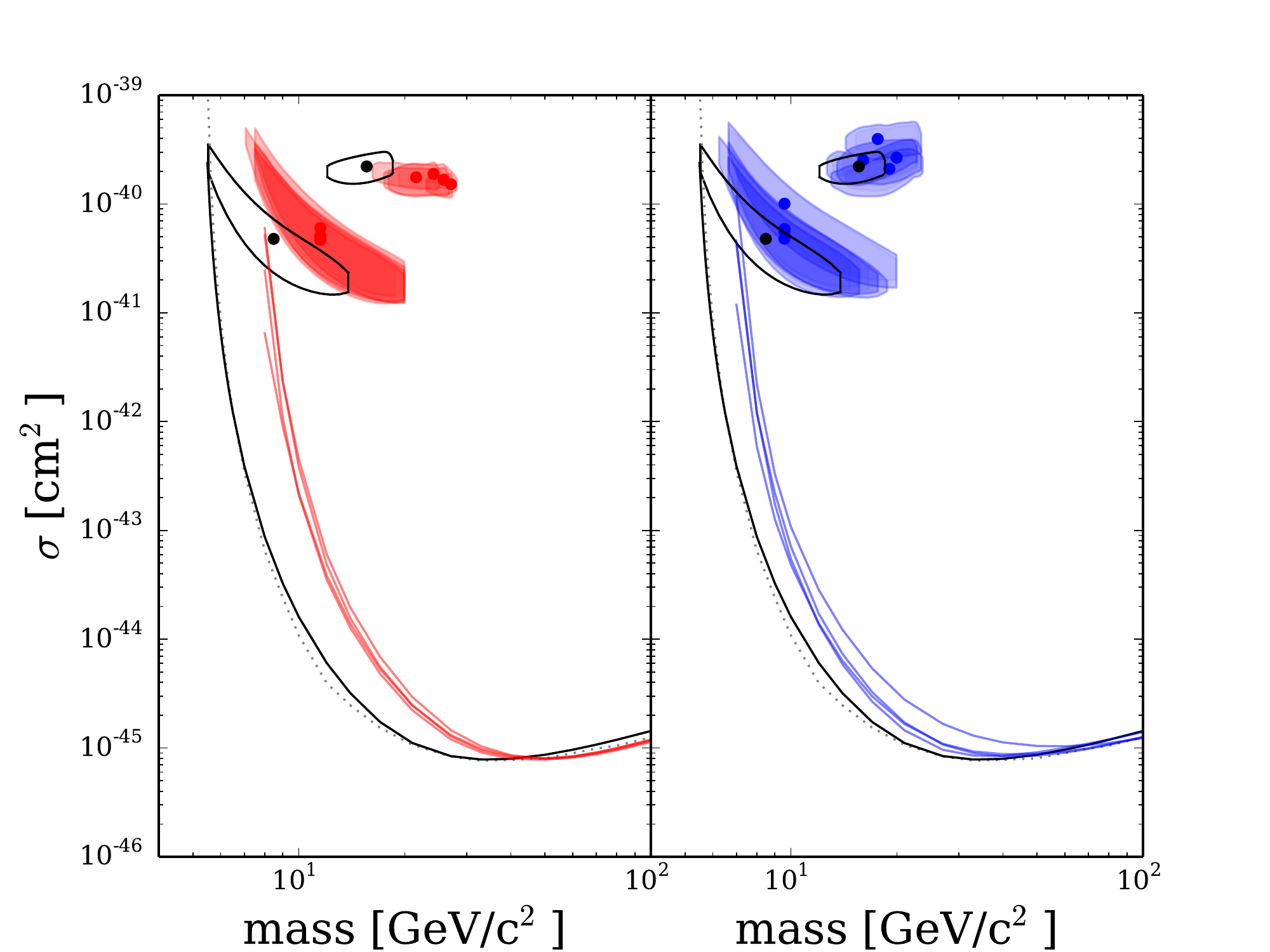}
\caption{Exclusion limits using various velocity distributions. Blue is from baryonic simulations, red from dark matter-only simulations, black is the SHM, and dotted black is the LUX result from \citet{FirstLUXResults}.  The large change expected from dark matter-only simulations (seen in the left-hand panel) is reduced  when baryons are included (right-hand panel).  The constraints on $\sigma$ remain tight at higher dark matter masses because there is less high velocity dark matter in the simulations than predicted by the SHM.  However, there is up to an order of magnitude change in the exclusion limits at low dark matter masses when using the simulated velocity distributions instead of the SHM.}
\label{sigma_m_LUX}
\end{figure*}

The critical quantity to compare for the direct detection search is $g(v_{\text{min}})$. In Figure \ref{g_earth} we show $g(v_{\text{min}})$ for the days of highest and lowest dark matter flux. It is clear that the baryonic runs and SHM have distributions more similar than between the SHM and the dark matter-only runs. When compared to the baryonic simulations, the SHM overpredicts the fraction of high-velocity dark matter, and underpredicts the low-velocity. Thus, the SHM will set conservative exclusion limits at high dark matter masses (which correspond to small $v_{\rm min}$), and overly optimistic bounds at low dark matter masses. However, for these masses the SHM would be a conservative choice for detection.
Though the baryonic runs are more similar to the SHM than the dark matter-only simulations are, deviations from the SHM are still present. This is consistent with expectations from previous work by \citet{Pillepich2014}.

Having introduced the connection between velocity distributions and the rate of events in direct detection experiments, we now calculate the experimental constraints on $m_\chi$ vs.~$\sigma$ one would obtain using our simulated distributions, rather than the SHM.
Our LUX analysis is less complex than that done by the LUX Collaboration \citep{FirstLUXResults} given our relative lack of knowledge about the background distributions. Instead we closely follow the procedure described in \citet{GreshamZurek2014} which was applied to results from the Xenon100 experiment \citep{XENON2012}. We use a maximum gap analysis as described in \citet{Yellin2002}, varying which events from the \citet{FirstLUXResults} dataset were included in our signal region to ensure our SHM limits agree with the experimental limits of \citet{FirstLUXResults}, which were obtained using the full instrument information. This required including two events: the events at $\sim 3$ and $19$ phe.
Both our CoGeNT and our DAMA/Libra analyses are chi-squared fits to the data presented in \citet{CoGeNT2013} and \citet{DAMALIBRA2010}, respectivley, using our velocity distributions and a Helm Form Factor \citep{Duda2007}.

We compare the annual modulation signal from the DAMA/Libra data to that inferred from our simulations. 
Unlike other experiments which seek to be ``zero-background,'' DAMA/Libra is sensitive to annual modulation in the rate of dark matter scattering, caused by the motion of the Earth into and out of the Galactic dark matter ``wind.'' As a result, the date when the Earth and Sun's combined motion is maximized is important for analyzing DAMA/Libra's data.
We have made a change to the day of peak flux used in favor of self consistency within the simulations, in addition to the change in the local standard of rest velocity (as used in our LUX analysis). The peak days used for the simulation processing are not those of Earth in its rotation; this is because the simulated dark matter population has a net direction after azimuthal averaging that is not purely tangential. This direction impacts the day of flux extrema by a few days for most of our simulated galaxies, and at most by a week.

Figure \ref{sigma_m_LUX} shows the effect of using our simulated velocity distributions on the DAMA/Libra and CoGeNT preferred $m_\chi - \sigma$ regions and the LUX $m_\chi - \sigma$ exclusion region. CoGeNT shows less scatter than DAMA/Libra as it is insensitive to the precise days of flux extrema. Importantly, the baryon simulations have a greater variation due to their more diverse velocity distributions. The precise origin of this diversity is left to a future paper, but the point we wish to emphasize in this work is that the large systematic shift indicated by the dark matter-only runs is not upheld by the baryonic runs. This is particularly true for CoGeNT and DAMA/Libra, but the baryonic simulations show a smaller shift even in the case of LUX.  

\begin{figure*}[!t]
\centering
\includegraphics[width=0.4\textwidth]{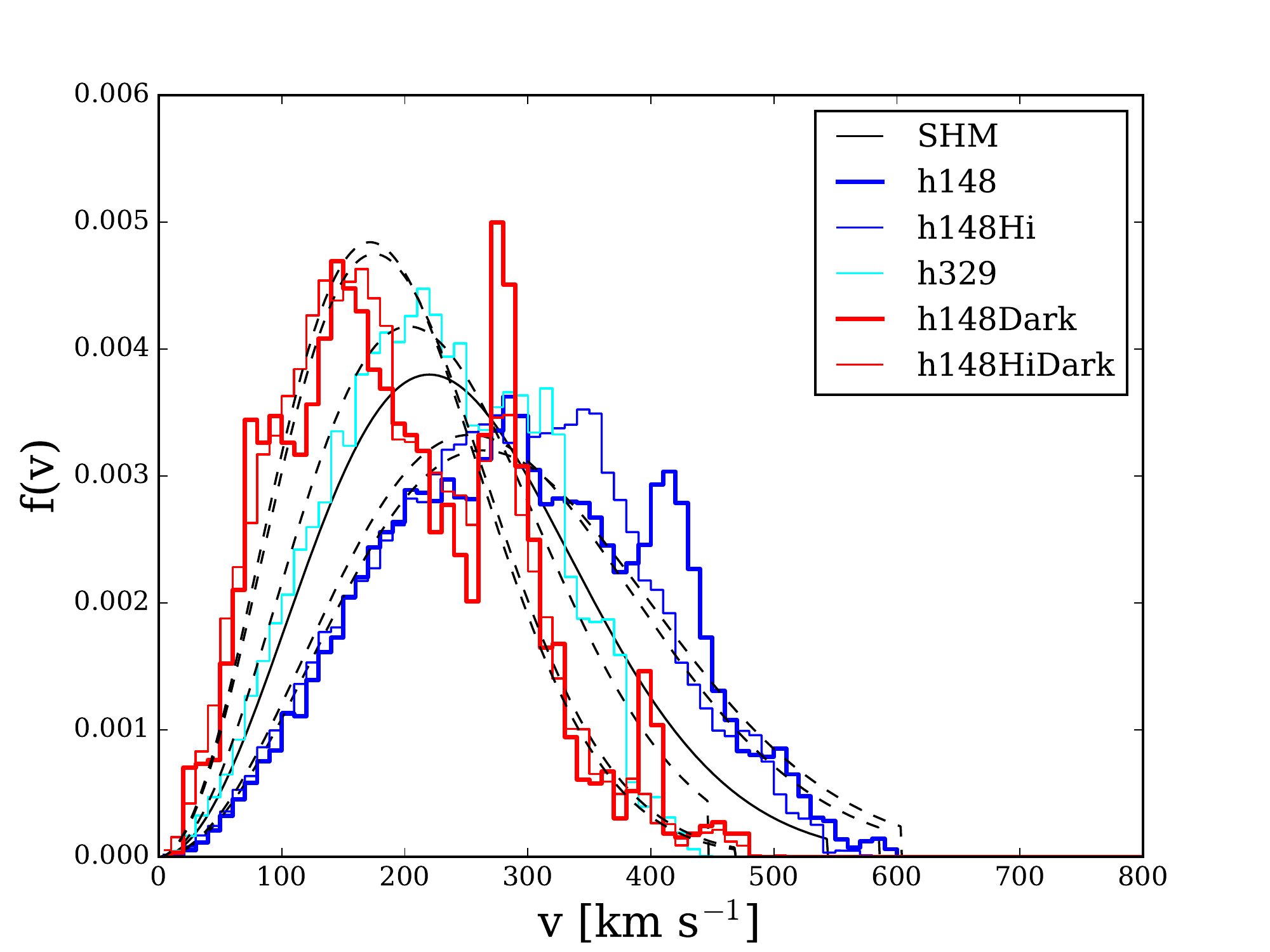}\includegraphics[width=0.4\textwidth]{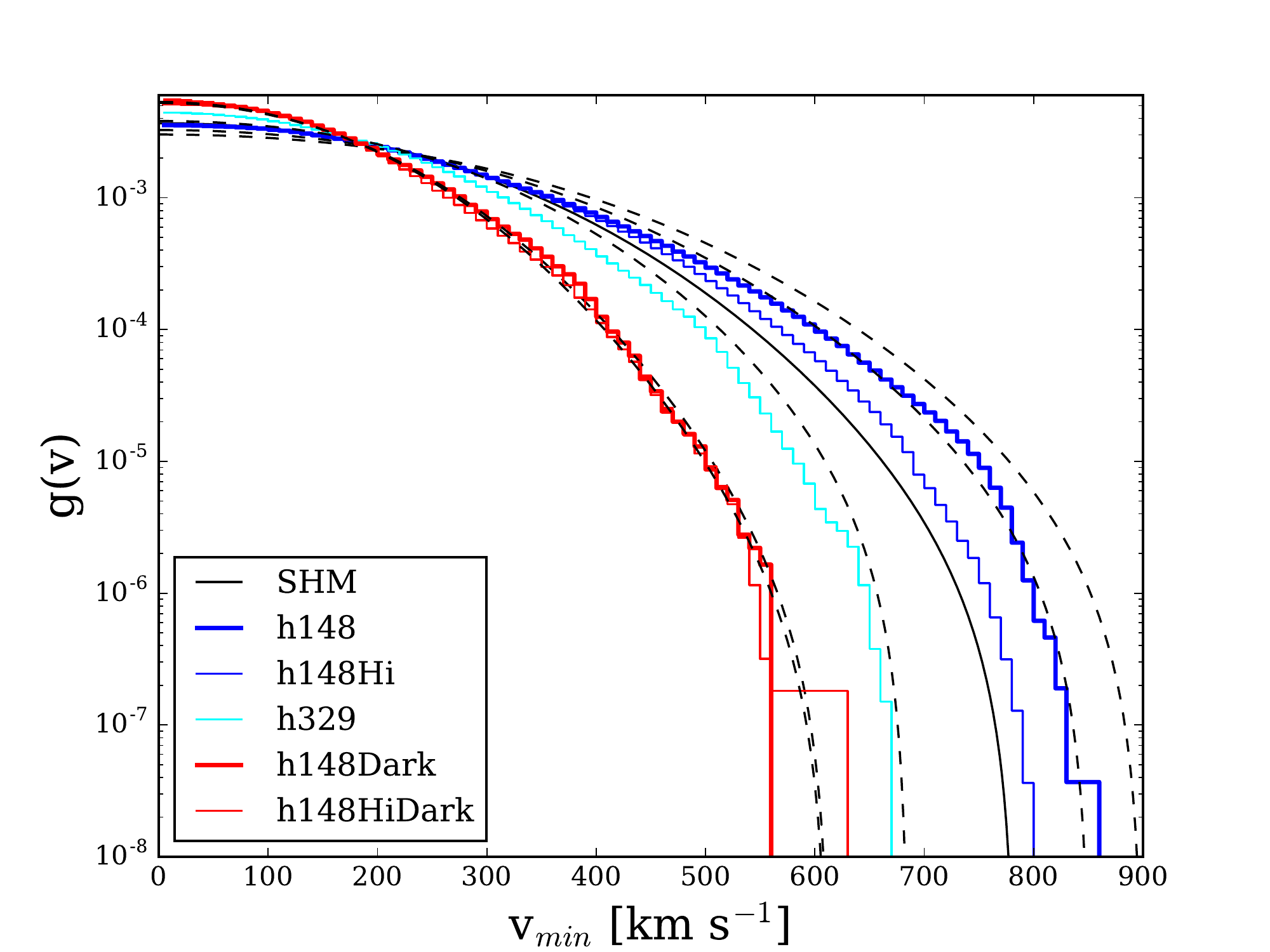}
\caption{Left: Local dark matter velocity distribution in the Galactic reference frame. The lower mass galaxy at our standard resolution, h329, is shown in cyan.  The higher mass galaxy, h148, is shown in blue at both higher resolution (solid) and the standard resolution (dashed). The corresponding dark matter-only runs of h148 are shown in red. The SHM is shown by the solid black line.  Best fit Maxwellians to each halo are shown by black dashed lines.  Relative to the SHM, only the lower mass h329 behaves similar to the previous galaxies we have examined, with a dearth of high velocity dark matter compared to the SHM.  The higher mass h148 has more high velocity material than the SHM.  However, the deficit of high velocity material still remains relative to a best fit Maxwellian.
Right: The weighted velocity distribution integral $g(v_{\rm min})$ in the Earth reference frame. Both h329 and the SHM have been boosted by 195 km s$^{-1}$, as in Figure~\ref{g_earth}.  However, h148 has been boosted by a Local Standard of Rest more appropriate for its mass, 240 km s$^{-1}$.}
\label{res_mass_tests}
\end{figure*}

Notably the strongest constraint on $\sigma$ occurs at higher dark matter mass, as there is less high velocity dark matter in the baryonic simulations than one would expect from the SHM. We can also see that the tension between the claimed signals and LUX's exclusion region is not alleviated, though we see up to an order of magnitude change in exclusion limit due to using our simulated velocity distributions instead of the SHM.

%%%%%%%%%%%%%%%%%%%%%%%%%%%%%%%%%%%%%%%%%%%%%%%%%%%%%%%%%%%%%%%%%%%%%%%%%%%%%%%

\section{Dependency on Resolution {, Mass Assembly History,} and Halo Mass}\label{Discussion}
In this section, we explore whether our conclusions depend on either resolution, {  mass accretion history,} or the mass of our halos.  As mentioned in Section \ref{Simulations}, we do this by investigating the velocity distribution of dark matter in two different simulations that span the range of masses allowed for the Milky Way.  These galaxies have different input physics and cosmology than the other four galaxies presented above, but they allow us to explore the role of halo mass, {  accretion history,} and resolution.

The left panel of Figure \ref{res_mass_tests} shows the local dark matter velocity distribution in the galactic reference frame for all versions of these galaxies compared to the SHM.  The lower mass galaxy, h329, shows similar trends to the four galaxies presented above, i.e., it displays a dearth of high velocity material relative to the SHM (solid black line).  For h148, only the dark matter-only versions of the run are missing high velocity material relative to the SHM.  The higher mass halo now shows an excess relative to the SHM for the baryonic runs.  However, we note that the baryonic runs still display a dearth of high velocity material relative to their best fit Maxwellians (dashed lines).  In other words, when you adopt a peak $v_0$ value more appropriate for these runs (254 km s$^{-1}$ and 230 km s$^{-1}$ for h148 and h148Hi, respectively, as seen in Table \ref{AppSimTable}), a Maxwellian fit still overpredicts the amount of high velocity material.

The right panel of Figure \ref{res_mass_tests} shows the weighted velocity distribution integral $g(v_{\rm min})$ in the Earth reference frame for these same halos.  As for the four galaxies presented above, the Local Standard of Rest in h329 is $\sim$195 km s$^{-1}$.  Hence, h329, its best fit Maxwellian, and the SHM have been boosted by 195 km s$^{-1}$, as in Figure~\ref{g_earth}.  However, the higher mass galaxy h148 has a larger Local Standard of Rest, $\sim$240 km s$^{-1}$.  Hence, all runs of h148 and their best fit Maxwellians were boosted by 240 km s$^{-1}$ instead.  It is now evident again that a Maxwellian predicts too much high velocity material within the Solar neighborhood of all of our simulated galaxies with baryons, independent of resolution or halo mass. 

\subsection{The Role of Halo Mass}

As expected, higher halo masses will lead to higher velocity dark matter being present in the solar neighborhood.  If halos across the range of masses allowed for the Milky Way \citep[$0.6 - 1.3\times10^{12}M_\odot$,][]{Kafle2012} are compared relative to a fixed Maxwellian like the SHM, the lowest mass end should fall below the SHM, while the highest mass end will exceed the SHM.  However, if instead each halo is fit with the Maxwellian most appropriate to it, we instead find that there is consistently less high velocity material than the Maxwellian predicts.

This interpretation is consistent with the results presented in \citet{Kelso:2016qqj}, where they used two galaxies that also spanned the range of allowed Milky Way halos masses.  \citet{Bozorgnia:2016ogo}, however, used halos that were much more massive.  Two of their high resolution {\sc apostle} simulations had masses at the upper end of the Milky Way's allowed range (1.64 and 2.15$\times$10$^{12}$ M$_{\odot}$), while the halos drawn from {\sc eagle} were more massive (2.76-14.25$\times$10$^{12}$ M$_{\odot}$).  The main impact of this much larger halo mass is that the local dark matter density is much higher, 0.4-0.7 GeV/cm$^3$ for the \citet{Bozorgnia:2016ogo} galaxies.  This caused their $\sigma_N$ -- $m_\chi$ limits to shift in a direction opposite to what we see in Figure \ref{sigma_m_LUX}, as their densities are higher than adopted by the SHM.  Our galaxies, on the other hand, have local dark matter densities $\sim$0.3 GeV/cm$^3$, consistent with what is commonly adopted by direct detection experiments when calculating the SHM.  For our galaxies, the shift in 
Figure \ref{sigma_m_LUX} with respect to the SHM is instead due to a lack of high velocity material in our simulated distributions.

\subsection{The Role of Mass Assembly History}\label{MAHSection}
 {
The four galaxies that form the main study in this paper use a WMAP3 cosmology, but the auxiliary halos discussed in this section use a Planck cosmology. The consistency in the deficit of high velocity particles indicates that the effect of the cosmological parameter change between WMAP3 and Planck is secondary for our purposes. However, Planck halos should, on average, have earlier formation times and higher concentrations than WMAP3 due to a higher $\sigma_8$ value in the Planck cosmology \citep{Dutton2014}.
}
Despite its Planck cosmology, h329 has a local dark matter density similar to all four of our WMAP3 galaxies.  h148, on the other hand, has a local dark matter density of 0.5 GeV/cm$^3$.  {  Comparison of the formation times in Table~\ref{AppSimTable} shows that the higher density in h148 is not due to an earlier formation.  In fact, h148 forms later than h329.}  Hence, the higher densities seem to be associated with higher mass halos, independent of cosmology. 

The change in local dark matter density seems to be the most significant effect of increasing halo mass.  That is, our halos across all masses, resolutions, and cosmology are missing high velocity material relative to their best fit Maxwellians, which will consistently shift our $\sigma_N$ -- $m_\chi$ as seen in Figure \ref{sigma_m_LUX} as long as a fixed value is adopted for the local dark matter density.  On the other hand, if the adopted local dark matter density is higher than adopted for the SHM, it instead shifts  {exclusion limits} in the opposite direction, as shown in \citet{Bozorgnia:2016ogo}.

 {
We also verified that merger history seems to be more important than mass assembly history in the resulting velocity distributions in the Solar neighborhood. 
From Table \ref{SimTable}, it can be seen that h277 and h239 have the most discrepant formation times.  However, they have similar velocity distributions in the Solar neighborhood at $z = 0$. Furthermore, h239 and h258 have similar mass assembly histories, but have very dissimilar velocity distributions in the Solar neighborhood. The specifics of the merger history seem more indicative of a galaxy's velocity structure than the overall mass assembly history.
}

\subsection{The Role of Resolution}

Our simulations are the highest resolution yet used to examine the velocity distribution of dark matter in the solar neighborhood.  Both \citet{Bozorgnia:2016ogo} and \citet{Kelso:2016qqj} have a force softening of $\sim$300 pc. \citet{Kelso:2016qqj} also found a hint that their velocity distributions were lower than a best fit Maxwellian, though they cautioned that their low resolution may lead to small sampling statistics.   They used similar numbers of dark matter particles to our original four galaxies, but with a much larger annulus (2-8 kpc rather than our 7.5-8.5 kpc annulus).  We confirm that the dearth of high velocity particles remains even at higher resolutions.  We note that h329 and h148 contain an additional factor of 4-10 times as many dark matter particles within our defined solar annulus as the four galaxies presented earlier, because the dark matter particles have been split even further to smaller masses.

The resolution of our simulations is comparable to \citet{Pillepich2014}, who also found a dearth of high velocity material in the Eris simulation relative to the SHM.  Hence, to date, all of the highest resolution simulations point to a lack of high velocity material in the Solar neighborhood relative to a Maxwellian.  Eris is also on the low mass end allowed for the Milky Way, but we have shown in this section that halo mass is not responsible for the lack of high velocity material.  The material is still missing, even at higher halos masses, as long as the comparison is made to a Maxwellian that is the best fit for the halo.

%%%%%%%%%%%%%%%%%%%%%%%%%%%%%%%%%%%%%%%%%%%%%%%%%%%%%%%%%%%%%%%%%%%%%%%%%%%%%%%

\section{Conclusions}\label{Conclusions}
We have compared the local velocity distribution in four high resolution cosmological simulations of Milky Way-analogue halos run both as dark matter-only and with baryons included. We find that the dark matter velocity distributions are influenced appreciably by the inclusion of baryons. We conclude that several important lessons can be drawn based on these results:

\begin{itemize}
\item{Due to the fact that they are less computationally expensive, dark matter-only simulations have previously been used to generate a statistical sample of Milky Way-mass galaxies in order to derive a functional form for the dark matter velocity distribution in the Solar neighborhood.  However, neglecting the impact of baryons will lead to erroneous results. As anticipated by \citet{Mao2013I}, the inclusion of baryons  leads to more high velocity dark matter in the Solar neighborhood. However, the shape of the velocity distribution function is also altered in ways not captured by functional forms fit to dark matter-only runs.  Ironically, we find that the inclusion of baryons leads to velocity distributions which bring direct detection experimental results closer to those extracted from the SHM than results obtained using dark matter-only simulations. 
In order to quantify how the range of velocity distributions depends on merger history or other factors, or to define a functional form that is better suited to the results seen in baryonic simulations, a larger sample of simulations will be needed to provide statistics. }

\item{For the purpose of interpreting direct detection experiments, it is critical to examine simulated velocity distributions \text{\it transformed} to the lab frame. The baryonic simulations can have significant velocity substructure that not only is missing in the dark matter-only runs, but that is neglected when derived in the Galactic reference frame, even if baryonic simulations are used. The results in the Galactic frame retain only information on the magnitude, not the directionality, of the dark matter velocities. The velocity space structure is vital to the direct detection search, and when the velocity distribution is made in the Galactic reference frame, the vector information is lost prematurely. Given the particle data afforded by a simulation, one should transform the velocities into the lab frame and \text{\it then} fit to a velocity distribution. }

\item{Our velocity functions extracted from baryonic simulations do not resolve the tension between the claimed positive signals of DAMA/Libra and CoGeNT and the null results from LUX.  However, our simulations predict that the SHM is conservative for the exclusion of dark matter when the bound is  dominated by the detector sensitivity to low velocity dark matter. 
If an experiment which is sensitive only to the high velocity tail of the dark matter distributions uses the SHM in its analysis, it would expect more dark matter interactions  than would be the case according to our simulations; in such an instance there could be a false rejection of the dark matter hypothesis. Mass ranges where experiments are dominated by scattering below about 200 km s$^{-1}$ will suffer from the opposite issue; here any signal would be amplified by using the SHM. }

\end{itemize}

In our sample of four simulations, we found greater diversity in the details of the velocity distributions in our baryonic runs than in our equivalent dark matter-only simulations. This indicates that the range of possible velocity distributions in realistic galaxies may be large, and highlights the need for large suites of high-resolution galaxy simulations including baryons in order to search for trends in velocity distribution as a function of galaxy merger history. While we need to consider the effects of baryons on dark matter, we also need to carefully choose halos to match the Milky Way's history and environment to be able to make specific statements about our local dark matter environment. 

It is again interesting to compare our results with the recent papers of \citealt{Bozorgnia:2016ogo} and \citealt{Kelso:2016qqj}. Though we compare a similar number of simulated galaxies with baryonic physics included, our galaxies have a deficit of high-velocity dark matter in the SPH simulations when compared to the SHM. This leads to our extrapolated direct detection limits being weaker than those predicted by the SHM. This is very different from the results of \citealt{Bozorgnia:2016ogo} and for one of the galaxies in \citet{Kelso:2016qqj}.  These authors instead find an increase in the high velocity distributions. However, in all cases, this increase is in high mass halos that have a larger density of dark matter in the Solar neighborhood.  The higher density seems to be the primary culprit responsible for the shift in the opposite direction to what we found for our galaxies in the $\sigma_N$ -- $m_\chi$ plane.  

Additionally, the resolution of our simulations ($\sim 170$~pc) is higher than that used in both \citealt{Bozorgnia:2016ogo} and \citealt{Kelso:2016qqj}. Resolution may play a critical role in the local dark matter density and the velocity distribution.  At high resolutions, energetic feedback from stars and supernova can transfer energy to the dark matter component, expanding the orbits of the dark matter.  \citet{Governato2010} showed that the energy transfer to the DM component from rapid gas outflows is drastically reduced at resolutions worse than 200-300pc, reducing any effect of baryonic physics on the dark matter structure. 

Overall, our results underline the need for a larger suite of high resolution simulations with baryonic physics that accurately satisfy a large number of observational constraints, in order to provide sufficient statistics for reliable extrapolation. In particular, the large variance within ``Milky Way-mass'' galaxies motivates efforts to find methods of relating the velocity structures found in simulations to observables in the Milky Way itself, in order to more accurately determine the velocity distribution not of an average Milky Way-like galaxy, but of the specific galaxy in which we reside.

\acknowledgements
Resources supporting this work were provided by the NASA High-End Computing (HEC) Program through the NASA Advanced Supercomputing (NAS) Division at Ames Research Center.  We thank Tom Quinn and James Wadsley for use of the proprietary code {\sc Gasoline}, and Charlotte Christensen for help in creating the simulations. The pynbody package \citep{pynbody} was used in portions of this analysis.

%%%%%%%%%%%%%%%%%%%%%%%%%%%%%%%%%%%%%%%%%%%%%%%%%%%%%%%%%%%%%%%%%%%%%%%%%%%%%%%

\newpage
\bibliographystyle{apj}
\bibliography{AstrophysicalUncertainties}

\end{document}